\documentclass[12pt]{article}
\usepackage{aaspp4}
\usepackage[final]{graphicx}
\graphicspath{{fig/}}

\begin{document}

\newcommand{\eg}{ e.g.}
\newcommand{\etal}{ et al.}
\newcommand{\nuc}[2]{\ensuremath{\mathrm {^{#2}#1}}}
\newcommand{\alp}{\ensuremath{\alpha}}
\newcommand{\edot}{\ensuremath {\dot \epsilon}}
\newcommand{\tnine}{\ensuremath{{T}_{9}}}
\newcommand{\tninei}{\ensuremath{{T}_{9i}}}
\newcommand{\rhoi}{\ensuremath{\rho_i}}
\newcommand{\ttt}[1]{\ensuremath{\times 10^{#1}}}
\newcommand{\mev}{\ensuremath{\mathrm{\thinspace MeV}}}
\newcommand{\erggs}{\ensuremath{\mathrm{\thinspace ergs \thinspace g^{-1} 
\thinspace s^{-1}}}}
\newcommand{\gcc}{\ensuremath{\mathrm{\thinspace g \thinspace cm^{-3}}}}
\newcommand{\kel}{\ensuremath{\mathrm{\thinspace K}}}
\newcommand{\calF} {\ensuremath {\mathcal {F}}}
\newcommand{\calR} {\ensuremath {\mathcal {R}}} 
\newcommand{\calRp}{\ensuremath {\mathcal {R'}}} 
\newcommand{\calG}{\ensuremath {\mathcal {G}}} 
\newcommand{\calM}{\ensuremath {\mathcal {M}}} 
\newcommand{\calZ}{\ensuremath {\mathcal {Z}}} 
\newcommand{\yf}{\ensuremath {\vec {Y}^{\calF}}} 
\newcommand{\yr}{\ensuremath {\vec {Y}^{\calR}}} 
\newcommand{\yrp}{\ensuremath {\vec {Y}^{\calRp}}} 
\newcommand{\yg}{\ensuremath {\vec {Y}^{\calG}}} 
\newcommand{\ygdot}{\ensuremath {\dot {{\vec Y}^{\calG}}}} 

\title{The QSE-reduced $\alpha$ Network}

\author{W. Raphael Hix}
\affil{Department of Astronomy, University of Texas, Austin, TX 78712}
\authoraddr{now at Joint Institute for Heavy Ion Research, Oak Ridge National 
Laboratory, Oak Ridge, TN 37831-6374}
\and
\author{Alexei M. Khokhlov}
\affil{Laboratory for Computational Physics and Fluid Dynamics, Code 6404, 
Naval Research Laboratory, Washington, DC 20735}
\and
\author{J. Craig Wheeler}
\affil{Department of Astronomy, University of Texas, Austin, TX 78712}
\and
\author{Friedrich-Karl Thielemann}
\affil{Department of Physics \& Astronomy, University of Basel, 
Klingelberstrasse 82, CH-4056 Basel Switzerland}

\begin{abstract}
Examination of the process of silicon burning, the burning stage that
leads to the production of the iron peak nuclei, reveals that the nuclear
evolution is dominated by large groups of nuclei in mutual equilibrium.
These quasi-equilibrium (QSE) groups form well in advance of the global 
Nuclear Statistical Equilibrium (NSE).  We present an improved ``minimal'' 
nuclear network, which takes advantage of quasi-equilibrium in order to 
calculate the nuclear evolution and energy generation while further reducing 
the computational cost compared to a conventional \alp-chain network.  During 
silicon burning, the resultant \emph{QSE-reduced} \alp\ network is twice as 
fast as the full \alp\ network it replaces and requires the tracking of only
half as many abundance variables, without significant loss of accuracy.  
When the QSE-reduced \alp\ network is used in combination with a conventional 
\alp\ network stretching from He to Si, the combined \alp 7 network provides an 
accurate approximation for all of the burning stages from He burning to NSE, 
while tracking only 7 abundances.  These reductions in computational cost and 
the number of species evolved make the \alp 7 network well suited for 
inclusion within hydrodynamic simulations, particularly those in 
multi-dimension.
\end{abstract}

\keywords{methods: numerical --- nuclear reactions, nucleosynthesis,
abundances --- stars: evolution --- supernovae: general}

\section{Introduction}

Nucleosynthesis calculations perform two functions in astrophysical models.
The most important function, in the context of the evolution of 
the encompassing hydrodynamic model, is the calculation of the rate of 
thermonuclear energy generation.  
Where nucleosynthesis occurs, it is often the dominant source of energy, 
making careful determination of this rate of energy generation vital to 
accurate hydrodynamic modeling.  The second function of a nucleosynthesis 
calculation is to determine the evolution of the nuclear species.  For
nuclear astrophysics and our understanding of the origin of the 
elements, this is of paramount importance.  Often the best comparisons of 
model calculations with observations also depend on detailed knowledge of 
the abundances of nuclear species.  Unfortunately, accurate calculation
of the nuclear evolution is computationally expensive.  For this reason,
it is common to include within a hydrodynamic model only a limited 
approximation for nuclear burning which estimates the rate of energy 
generation.  The detailed nuclear evolution is then calculated at a later time 
using the thermodynamic trajectories from the hydrodynamic model.  Such a 
scheme is referred to as post-processing nucleosynthesis.  The success of 
such a post-processing scheme depends largely on the accuracy with which the 
limited approximation which is included within the dynamical calculation
can estimate the rate of energy generation.  
Frequently this inline nuclear burning calculation is a small network, trading 
a limited ability to follow the nuclear evolution for speed. 

Tracking the nuclear evolution from helium burning through to NSE requires 
a network that includes nuclei from \alp-particles to Zn.  Silicon burning 
presents a particular problem as material proceeds from silicon to the iron 
peak not via heavy ion captures but through a chain of photodisintegrations 
and light particle captures.  The minimal nuclear set which can follow this 
evolution is the set of \alp-particle nuclei; \alp, \nuc{C}{12}, \nuc{O}{16}, 
\nuc{Ne}{20}, \nuc{Mg}{24}, \nuc{Si}{28}, \nuc{S}{32}, \nuc{Ar}{36}, 
\nuc{Ca}{40}, \nuc{Ti}{44}, \nuc{Cr}{48}, \nuc{Fe}{52}, \nuc{Ni}{56}, 
\nuc{Zn}{60}.  
For convenience we will label this full set \calF\ and refer to its abundance 
as \yf.  Silicon burning in fact presents a larger problem, as the nuclear 
flow from silicon to the iron peak nuclei does not generally procede through 
nuclei with N=Z, especially when significant neutronization has occurred 
(\cite{HiTh96} : henceforth HT96).  In some models, however, such 
compromise is made necessary by the computational limitations.

The nature of nuclear network calculations has been extensively reviewed 
elsewhere (see, \eg, \cite{Woos86}, \cite{ThNH94} or \cite{Arne96}), 
so we will discuss it only briefly here.  From a set of nuclear abundances, 
and the reaction rates for the 
reactions which link them, one can calculate the time derivatives of the 
abundances, $\dot {\vec Y}$.  With these derivatives, the abundances of the 
included nuclei, at time $t$, and over timestep $\Delta t$, are evolved 
according to the finite difference prescription
\begin{equation}
{{\vec Y(t+\Delta t)- \vec Y(t)} \over {\Delta t}} =
\begin{array}{ll}
\dot {\vec Y}(t) & (explicit) \\
\dot {\vec Y}(t+\Delta t) & (implicit)\ . \\
\end{array} 
\label{eq:deriv}
\end{equation}
For the stiff set of non-linear differential equations which form most 
nuclear networks, a fully implicit treatment is most successful.  Solving 
the implicit version of Eq.~\ref{eq:deriv} is equivalent to finding the 
zeros of the set of equations
\begin{equation}
\vec \calZ(t+\Delta t)\equiv {{\vec Y(t+\Delta t)- \vec Y(t)} \over {\Delta 
t}} - \dot {\vec Y}(t+\Delta t) =0 \ .
\end{equation}
This is done using the Newton-Raphson method (see, \eg, \cite{NumRec}), 
which is based on the Taylor series expansion of 
$\vec \calZ(t+\Delta t)$, with the trial change in abundances given by
\begin{equation}
\Delta \vec Y = \left( \partial \vec \calZ (t+\Delta t) \over \partial \vec Y 
(t+\Delta t) \right)^{-1} \vec \calZ \ ,
\label{eq:dely}
\end{equation}
where $\partial \vec \calZ / \partial \vec Y $ is the Jacobian of $\vec 
\calZ$.  Iteration continues until $\vec Y(t+\Delta t)$ converges.

The conclusion of silicon burning with nuclear statistical equilibrium (NSE) 
offers a hint to a more efficient means of evolving nuclear abundances, 
since the equilibrium distribution reduces the number of independant variables 
which must be tracked.  NSE grows from several groups of nuclei which form 
local equilibria.  The existence of these quasi-equilibrium groups frees one 
from accounting for the changes in the abundances of each of the members of 
these groups, as the changes within the groups can be accounted for by the 
changes of a few crucial abundances.  Previous authors, most notably 
\cite{BoCF68}: henceforth BCF), have attempted to use quasi-equilibrium to 
simplify the calculation of silicon burning.  Having noted the computational 
difficulties encountered by \cite{TrCG66} in their pioneering nuclear network 
study of silicon burning, BCF 
sought to model silicon burning with a single quasi-equilibrium group, 
stretching from \nuc{Mg}{24} and \nuc{Si}{28} through the iron peak nuclei.  
BCF postulated that the downward flow from this group occurred via $\rm 
^{24}Mg (\gamma,\alpha)^{20}Ne$ and that the flow of $\alpha$-particles 
downward through \nuc{Ne}{20}, \nuc{O}{16}, and \nuc{C}{12} conserved the 
abundances of these nuclei.  From these assumptions, they derived expressions 
for the abundances of these light $\alpha$-particle nuclei and for the downward 
flux of $\alpha$-particles which depended only on the abundances of 
$\alpha$-particles and \nuc{Si}{28}.  Though this model was moderately 
successful in replicating the abundance pattern found in the solar system from 
A=30 to A=60, it had clear limitations when compared to the network 
calculations of TCG, particularly at early times when the assumption of 
conservative flow was most suspect.  \cite{WoAC73}: henceforth WAC) concluded 
that such efforts were at best marginally successful, in part because as 
matter cools the photodisintegration reactions freeze out well before the 
particle captures cease, thereby destroying quasi-equilibrium.  WAC found a 
further complication in that, at early times, their models exhibited two 
separate QSE groups, one around silicon and the other containing the iron peak 
nuclei.  WAC did demonstrate that, with the assumption of QSE, reasonable 
estimates of the time required to merge the two QSE groups and the time to 
reach silicon exhaustion could be calculated.  HT96 demonstrated additional complications as both the path of the nuclear 
flow from the silicon group to the iron peak group and the membership of these 
groups depended strongly on the amount of neutronization.  HT96 further 
demonstrated that for material which has undergone significant neutronization, 
the separation between the silicon and iron peak groups persists through a 
significant part of the nuclear evolution.  In spite of these complications, 
\cite{HiTh98}: henceforth HT98) showed that QSE could provide a reasonable 
estimate of the abundances of many species during silicon burning, even under 
explosive conditions, where the photodisintegration reactions freeze out 
before their corresponding particle captures.  \cite{Arne96} provides a 
more complete history of the use of QSE.

\section{The QSE-reduced Network}

The objective of the QSE-reduced \alp\ network is to evolve \yf\ during silicon 
burning, and calculate the resulting energy generation, in a more efficient 
way.  Under conditions where QSE applies, the existence of the silicon and 
iron peak QSE groups allows calculation of the abundances of these 14 nuclei 
from a reduced set of 7 abundances.  For the members 
of the silicon group (\nuc{Si}{28}, \nuc{S}{32}, \nuc{Ar}{36}, \nuc{Ca}{40}, 
\nuc{Ti}{44}), the individual abundances can be calculated from the
assumption of QSE by
\begin{equation}
Y_{QSE,\mathrm{Si}}(^AZ)={{C(^AZ)}\over{C(\nuc{Si}{28})}} Y(\nuc{Si}{28}) 
Y_{\alpha}^{{A-28} \over 4} \ ,
\label{eq:yqsi}
\end{equation}
where we have defined
\begin{equation}
C(^AZ)= {G(^AZ) \over 2^A} {\left(\rho N_A \over \theta \right)}^{A-1} 
A^{3 \over 2} \exp {\left( B(^AZ) \over {k_B T} \right)} \ ,
\end{equation}
for later convenience and 
\begin{displaymath}
\theta = \left( {m_u k_B T \over 2 \pi \hbar^2 } \right)^{3/2} \ . 
\end{displaymath}
The functions $G(^AZ)$ and $B(^AZ)$ are the partition function 
and binding energy of the nucleus $^AZ$, $N_A$ is Avagadro's number, $k_B$ is 
Boltzmann's constant, and $\rho$ and $T$ are the density and temperature of 
the plasma.  $Y_{\alpha}$ is the abundance of free \alp-particles and the 
integer $(A-28)/4$ is the number of \alp-particles needed to construct 
$^{A}Z$ from \nuc{Si}{28}.  Similarly, the abundances of the members of the 
iron peak group (\nuc{Cr}{48}, \nuc{Fe}{52}, \nuc{Ni}{56}, \nuc{Zn}{60}) 
can be calculated by
\begin{equation}
Y_{QSE,\mathrm{Ni}}(^AZ)={{C(^AZ)}\over{C(\nuc{Ni}{56})}} Y(\nuc{Ni}{56}) 
Y_{\alpha}^{{A-56} \over 4} \ .
\label{eq:yqfe}
\end{equation}
Thus, under conditions where QSE applies, \yf\ can be expressed as a 
function of the abundances of the reduced nuclear set \calR, defined as 
[\alp, \nuc{C}{12}, \nuc{O}{16}, \nuc{Ne}{20}, \nuc{Mg}{24}, \nuc{Si}{28}, 
\nuc{Ni}{56}], and it is therefore sufficient to evolve only \yr.  It 
should be noted that WAC and HT96 have shown that \nuc{Mg}{24} is 
ordinarily a member of the silicon QSE group during silicon burning, 
leaving only 6 independant abundances, but for ease in integration with the 
conventional nuclear network which we will discuss in \S 4, we will evolve 
the \nuc{Mg}{24} abundance independantly (see \cite{Arne96}).

While \yr\ is a convenient set of abundances for the calculation of \yf, it is 
not the most efficient set to evolve, primarily because of the complexity of 
calculating the necessary time derivatives and their Jacobian.  It is 
much easier to calculate the time derivatives of each QSE group as a 
whole.  To this end, we define \yg = [$Y_{\alpha G}$, Y(\nuc{C}{12}), 
Y(\nuc{O}{16}), Y(\nuc{Ne}{20}), Y(\nuc{Mg}{24}), $Y_{SiG}$, $Y_{FeG}$], where
\begin{eqnarray}
Y_{\alpha G} & = & Y_{\alpha} + \sum_{i \in Si \ group} {{A_{i}-28} \over 4} 
Y_{i} + \sum_{i \in Fe \ group} {{A_{i}-56} \over 4} Y_{i} \ , \nonumber \\
Y_{Si G} & = & \sum_{i \in Si \ group} Y_{i} \ , \label{eq:yg} \\
Y_{Fe G} & = & \sum_{i \in Fe \ group} Y_{i} \ . \nonumber 
\end{eqnarray}
Physically, $Y_{Si G}$ and $Y_{Fe G}$ represent the total abundances of the 
silicon and iron peak QSE groups, while $Y_{\alpha G}$ represents the sum of 
the abundances of free \alp-particles and those \alp-particles required to 
build the members of the QSE groups from \nuc{Si}{28} or \nuc{Ni}{56}.

Employing the group abundance set \calG\ also reduces the number of reactions 
whose flux must be calculated since quasi-equilibrium allows one to ignore the 
reactions among the members of the QSE groups.  Unfortunately, the rates for 
this reduced set of reactions are still functions of \yf.  For example, the 
nuclear flux into the iron peak group, 
\begin{equation}
\dot Y_{Fe G} = \rho \langle \sigma v \rangle Y_{\alpha} Y(\nuc{Ti}{44}) - 
\lambda Y(\nuc{Cr}{48}) \ ,
\label{eq:dyfegf}
\end{equation}
depends not on the group abundances, $Y_{\alpha G}$, $Y_{Si G}$ and 
$Y_{Fe G}$, but on the abundances of free \alp-particles, \nuc{Ti}{44}, 
and \nuc{Cr}{48}.  
Since these time derivatives are not easily expressed in terms of \yg,  one 
must calculate  \ygdot\ from \yf\ or alternately, using Eqs.~\ref{eq:yqsi} 
and \ref{eq:yqfe}, from \yr.  Applying Eqs.~\ref{eq:yqsi} and \ref{eq:yqfe} to 
Eq.~\ref{eq:dyfegf} results in
\begin{equation}
\dot Y_{Fe G} = \rho \langle \sigma v \rangle {C(\nuc{Ti}{44}) \over C(\nuc{Si}{28})}
Y(\nuc{Si}{28}) Y_{\alpha}^{5} - \lambda {C(\nuc{Cr}{48}) \over 
C(\nuc{Ni}{56})} Y(\nuc{Ni}{56}) Y_{\alpha}^{-2} \ .
\label{eq:dyfegr}
\end{equation} 
Thus, to evolve \yg\ over time, one must repeatedly solve for \yr.  By applying 
Eqs.~\ref{eq:yqsi} and \ref{eq:yqfe} to Eq.~\ref{eq:yg}, one may write \yg\ as 
a function of \yr, with seven independant equations for the seven abundances 
of \yr\ implying a unique relation between \yg\ and \yr.  Fortunately, with 
the exception of $Y_{\alp}$, the relations between \yg\ and \yr\ are linear.
  
Further complicating the calculation, however, is the need in Eq.~\ref{eq:dely} 
for the Jacobian of $\vec \calZ$.  This requires knowledge of $\partial 
\ygdot/\partial \yg$, which can not be calculated directly since \ygdot\ can 
not be expressed in terms of \yg.  Instead, we have been successful using the 
chain rule,
\begin{equation}
{{\partial \ygdot} \over {\partial \yg}} = {{\partial \ygdot} \over {\partial 
\yr}} {{\partial \yr} \over {\partial \yg}}
\label{eq:chain}
\end{equation}
to calculate the Jacobian.  Analytically, the first term on the righthand 
side of Eq.~\ref{eq:chain} is easily calculated from the sums of reaction 
terms.  For example, from Eq.~\ref{eq:dyfegr} we can see that the non-zero 
terms of $\partial \dot {Y}_{Fe G}/\partial \yr$ are 
\begin{eqnarray}
{{\partial \dot {Y}_{Fe G}} \over {\partial Y_{\alpha}}}  & = & 
5 \langle \sigma v \rangle {C(\nuc{Ti}{44}) \over C(\nuc{Si}{28})}
Y(\nuc{Si}{28}) Y_{\alpha}^{4} + 2 \lambda {C(\nuc{Cr}{48}) \over 
C(\nuc{Ni}{56})} Y(\nuc{Ni}{56}) Y_{\alpha}^{-3} \ ,  \nonumber \\
{{\partial \dot {Y}_{Fe G}} \over {\partial Y(\nuc{Si}{28})}}  & = &  \langle 
\sigma v \rangle {C(\nuc{Ti}{44}) \over C(\nuc{Si}{28})} Y_{\alpha}^{5} 
\quad \mathrm{and} \quad 
{{\partial \dot {Y}_{Fe G}} \over {\partial Y(\nuc{Ni}{56})}}  = 
\lambda {C(\nuc{Cr}{48}) \over C(\nuc{Ni}{56})} Y_{\alpha}^{-2} \ . 
\label{eq:term1} 
\end{eqnarray}

Calculation of the second term on the righthand side of Eq.~\ref{eq:chain} 
is more complicated, requiring implicit differentiation of the definition 
of \yg\ with respect to \yg, using Eq.~\ref{eq:yg}.  As an example, in order 
to calculate $\partial \dot {Y}_{Fe G}/ \partial Y_{Fe G}$, we differentiate 
Eq.~\ref{eq:yg} with respect to $Y_{Fe G}$, resulting in 3 equations for 
three unknowns, $\partial Y_{\alpha} / \partial Y_{Fe G}$, $\partial 
Y(\nuc{Si}{28}) / \partial Y_{Fe G}$, and $\partial Y(\nuc{Ni}{56}) / \partial 
Y_{Fe G}$, while the other 4 terms of $\partial \yr / \partial Y_{Fe G}$ 
are manifestly zero.  Solving these 3 equations results in the 3 terms 
\begin{eqnarray}
{\partial Y_{\alpha} \over \partial Y_{Fe G}} &=& -{\calM_{Fe}^{1} \over 
\calM_{Fe}^{0}} {\calM_{\alp}}^{-1} \nonumber \\
{\partial Y(\nuc{Si}{28}) \over \partial Y_{Fe G}} &=& {Y(\nuc{Si}{28}) \over 
Y_{\alp}} {\calM_{Si}^{1} \over \calM_{Si}^{0}} {\calM_{Fe}^{1} \over 
\calM_{Fe}^{0}} {\calM_{\alp}}^{-1} \label{eq:term2} \\
{\partial Y(\nuc{Ni}{56}) \over \partial Y_{Fe G}} &=& {Y(\nuc{Ni}{56}) \over
\calM_{Fe}^{0}} \left( 1+ {{\calM_{Fe}^{1}}^{2} \over \calM_{Fe}^{0}}
{{\calM_{\alp}}^{-1} \over Y_{\alpha}} \right) \ .  \nonumber
\end{eqnarray}
where we have used the following definitions for simplification
\begin{eqnarray}
\calM_{Si}^j&=&\sum_{i \in Si\ Group} \left({A-28} \over 4\right)^j Y_i \ , 
\qquad 
\calM_{Fe}^j=\sum_{i \in Fe\ Group} \left({A-56} \over 4\right)^j Y_i 
\nonumber \\
\calM_{\alp} &=& 1+{1 \over Y_{\alp}} \left( \calM_{Si}^{2} + \calM_{Fe}^{2} 
- {{\calM_{Si}^{1}}^{2} \over \calM_{Si}^{0}} - {{\calM_{Fe}^{1}}^{2} \over 
\calM_{Fe}^{0}} \right) \ . \label{eq:M}
\end{eqnarray}
Multiplying each term in Eq.~\ref{eq:term2} by the respective component of 
Eq.~\ref{eq:term1} and summing produces $\partial \dot {Y}_{Fe G}/ \partial 
Y_{Fe G}$.

\section{Silicon burning with the QSE-reduced \alp\ network}

\begin{figure}[tbp]
    \centering 
    \includegraphics[angle=90,width=\textwidth]{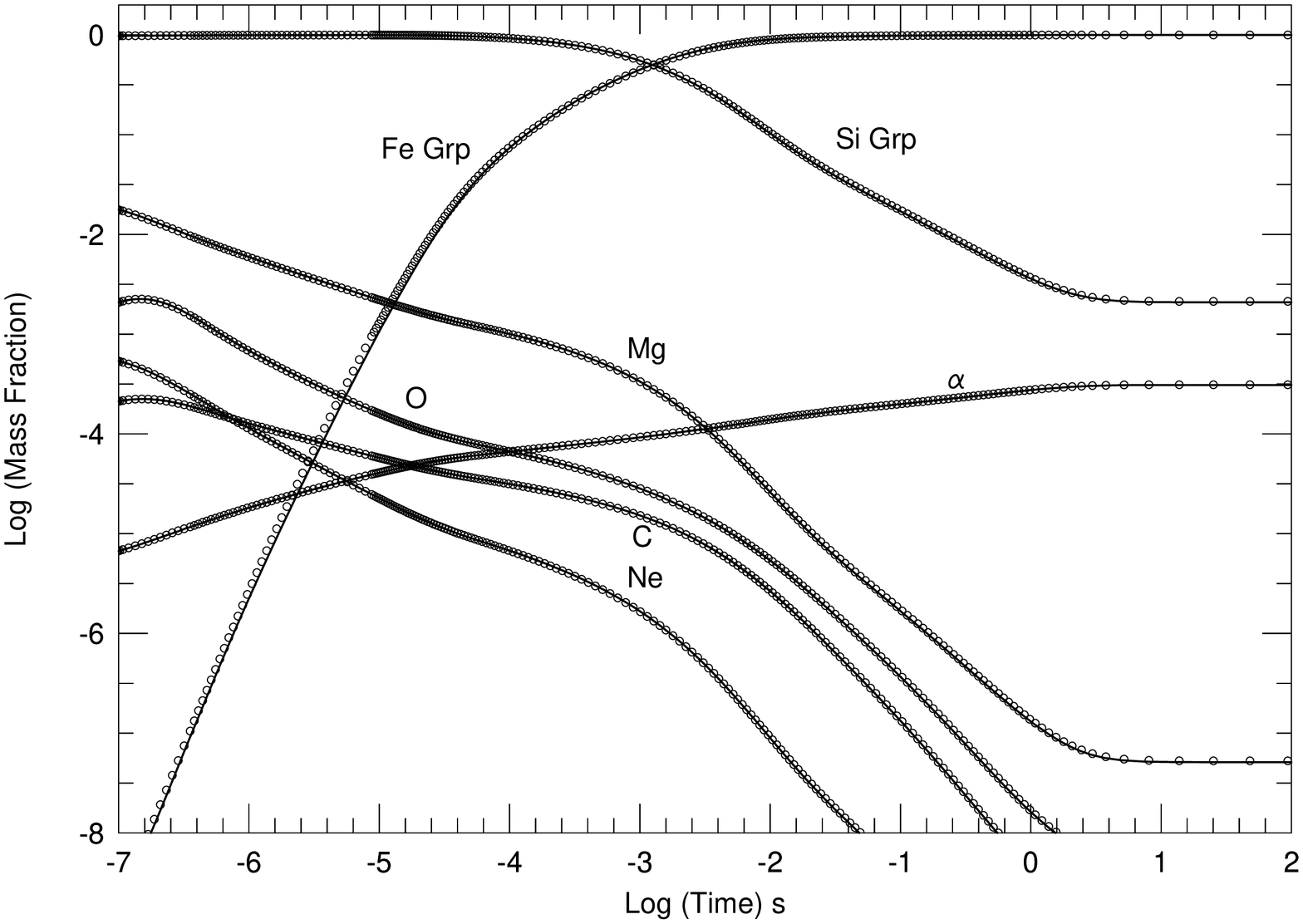} 
    \caption{Evolution of the independant nuclear mass fractions for constant 
    thermodynamic conditions, \tnine=5 and $\rho= 10^{9} \gcc$.  The solid 
    lines display the evolution due to a conventional \alp\ network, the 
    circles show the evolution by the QSE-reduced \alp\ network. The Silicon 
    group mass fraction is the sum of the mass fractions of \nuc{Si}{28}, 
    \nuc{S}{32}, \nuc{Ar}{36}, \nuc{Ca}{40}, and \nuc{Ti}{44}.  The Iron 
    group mass fraction is the sum of the mass fractions of \nuc{Cr}{48}, 
    \nuc{Fe}{52}, \nuc{Ni}{56}, and \nuc{Zn}{60}}
    \label{fig:xc59}	
\end{figure}

In this section we will demonstrate the accuracy with which the QSE-reduced 
\alp\ network duplicates the results of the full 14 element \alp\ network
for silicon burning.  Our first examples are nucleosynthesis calculations 
occuring under constant temperature and density.  While these calculations provide the 
least challenging comparison, they also allow comparison with NSE, which
should represent the final abundances of these calculations. 
Figure~\ref{fig:xc59} offers comparison of the mass fractions of the 7 
independant species; \alp-particles, \nuc{C}{12}, \nuc{O}{16}, \nuc{Ne}{20}, 
\nuc{Mg}{24}, and the silicon and iron peak groups, as evolved by the
QSE-reduced and conventional \alp\ networks for silicon burning at $5 \ttt{9} 
\kel$ and a density of $10^{9} \gcc$.  Apart from an early enhancement by
the QSE-reduced network of the iron peak mass fraction (20\% after $10^{-6}$
seconds), these mass fractions typically agree to within 1\%.  Since the 
nuclear energy release depends linearly on the abundance changes, differences 
in small abundances have little effect on the nuclear energy store.  In 
this case, the difference in the rate of energy generation calculated by
the two networks is $<1\%$ at $10^{-6}$ and $10^{-4}$ seconds.  This
difference is significantly smaller than the variation, shown by both
networks, in the rate of energy generation between timesteps, with \edot\ 
typically declining by 5\% per timestep over this interval.  

\begin{table}
    \centering 
    \caption{Comparison between the QSE-reduced and conventional network 
    abundances after 1.06 \ttt{-4} seconds with \tnine=5 and $\rho=10^9 \gcc$}
    \label{tab:const1}
    \vspace{12pt}
    \begin{tabular}{c|ll}
        \tableline
        Nucleus & \multicolumn{1}{c}{$Y_{net}$}&\multicolumn{1}{c}{$Y_{qse}$}\\
        \tableline
        \nuc{He}{4} & 1.67\ttt{-5} & 1.68\ttt{-5} \\ 
        \nuc{C}{12} & 2.59\ttt{-6} & 2.58\ttt{-6} \\
        \nuc{O}{16} & 4.09\ttt{-6} & 4.09\ttt{-6} \\
        \nuc{Ne}{20}& 3.29\ttt{-7} & 3.28\ttt{-7} \\
        \nuc{Mg}{24}& 4.11\ttt{-5} & 4.11\ttt{-9} \\
        \nuc{Si}{28}& 1.88\ttt{-2} & 1.89\ttt{-2} \\
        \nuc{S}{32} & 8.40\ttt{-3} & 8.36\ttt{-3} \\
        \nuc{Ar}{36}& 2.07\ttt{-3} & 2.04\ttt{-3} \\
        \nuc{Ca}{40}& 1.27\ttt{-3} & 1.24\ttt{-3} \\
        \nuc{Ti}{44}& 1.52\ttt{-5} & 1.48\ttt{-6} \\
        \nuc{Cr}{48}& 5.63\ttt{-5} & 5.45\ttt{-5} \\
        \nuc{Fe}{52}& 2.67\ttt{-4} & 2.69\ttt{-4} \\
        \nuc{Ni}{56}& 1.11\ttt{-3} & 1.13\ttt{-3} \\
        \nuc{Zn}{60}& 6.65\ttt{-8} & 6.73\ttt{-8} \\
        \tableline
    \end{tabular}
\end{table}

With such good agreement between the respective abundances of free 
\alp-particles and the two QSE group abundances, significant variations in 
abundance among the individual members of the QSE groups can only result 
from deviations from QSE. At early times, the small abundances within the 
iron peak reduce the accuracy of QSE at predicting the individual 
abundances of members of the iron peak group.  Much of the enhanced mass 
fraction of the iron peak nuclei at early times, seen in 
Fig.~\ref{fig:xc59}, is due to the QSE-reduced network's emphasis on 
heavier nuclei at the expense of \nuc{Cr}{48}.  After an elapsed time of 
$10^{-6}$ seconds, the average mass of the iron peak nuclei, $\bar 
A_{FeG}$, is 49.2 according to the conventional network and 52.6 according 
to the QSE-reduced network.  As a result, the abundances of \nuc{Cr}{48} 
and \nuc{Fe}{52} calculated by the QSE-reduced network are 38\% and 164\% 
of their conventional network values, while \nuc{Ni}{56} and \nuc{Zn}{60} 
are 16 times more abundant than the conventional network predicts.  As the 
iron peak nuclei become more abundant, QSE provides a better estimate of 
the relative abundances within the group, reducing such discrepancies.  By 
the time the iron peak nuclei represent a significant portion of the mass, 
the differences in the abundance predictions for all nuclei are only a few 
per cent.  Comparison of columns 2 and 3 of Table~\ref{tab:const1} show 
that after an elapsed time of $10^{-4}$ seconds, \nuc{Cr}{48} displays the 
largest difference, with the QSE-reduced network's abundance equal to 97\% 
of the abundance calculated by the conventional network.  In the context of 
the network evolution, this small difference is less than half of the 
abundance change for \nuc{Cr}{48} that occurs over the following timestep.  
As each network reaches its respective equilibria the abundance predictions 
of these networks continue to differ by at most 3\%, even among the nuclei 
with the smallest abundances.  Not surprisingly, in view of these small 
abundance differences, the difference in the total energy released by these 
networks is less than .1\%.  Comparison of the network abundances with 
abundances calculated from NSE show a similarly low level of difference.  
It should be noted that these NSE abundances, calculated assuming that the 
\alp\ nuclei are the only available states in the equilibrium, differ 
noticably from NSE calculations which include all nuclei, even for 
$Y_{e}=0.5$.

\begin{figure}[tbp]
	\centering
	\includegraphics[angle=90,width=\textwidth]{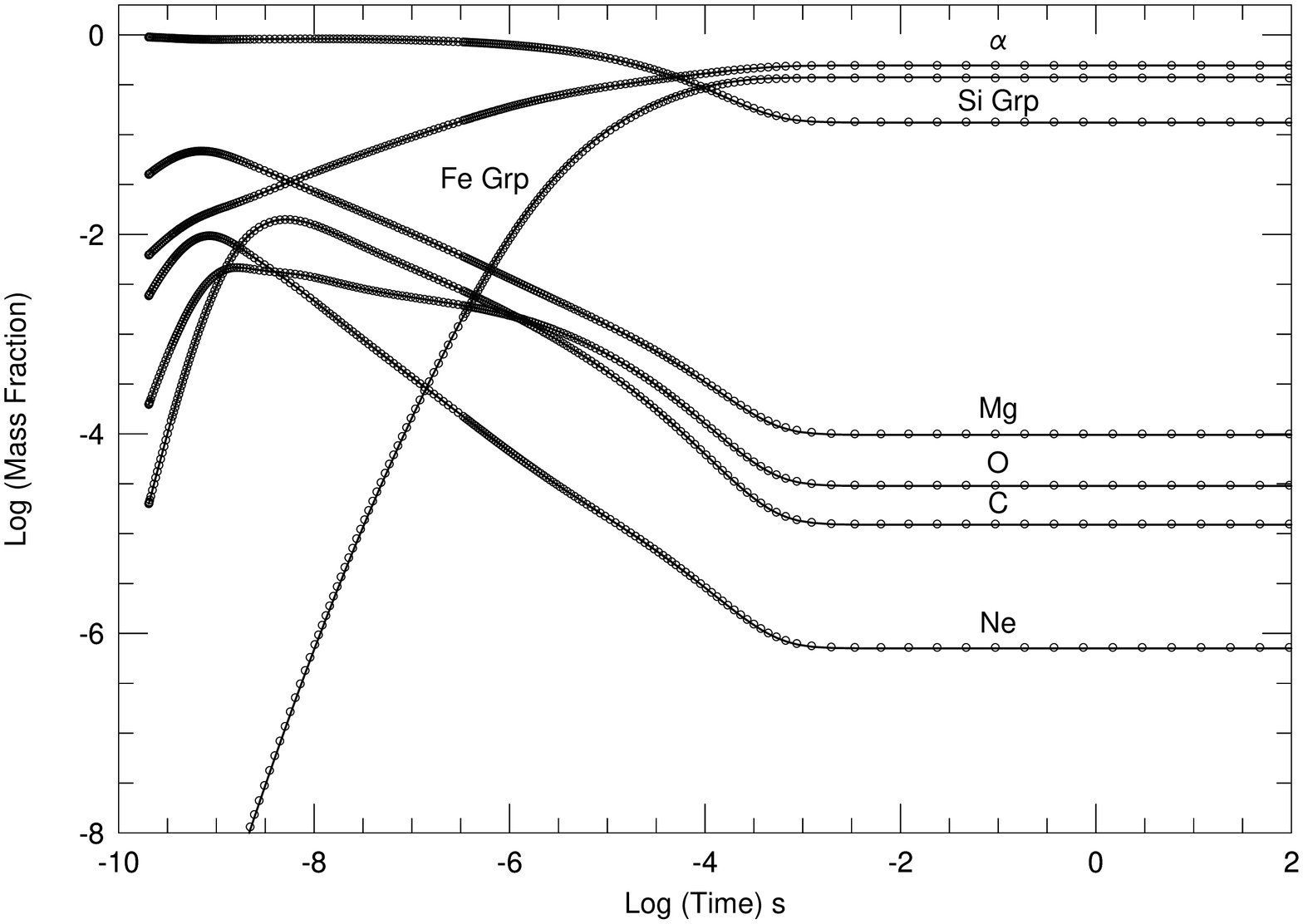} 
    \caption{Evolution of the independant nuclear mass fractions for constant 
    thermodynamic conditions, \tnine=6 and $\rho= 10^{7} \gcc$.  The solid 
    lines display the evolution due to a conventional \alp\ network, the 
    circles show the evolution by the QSE-reduced \alp\ network.  The Silicon 
    group mass fraction is the sum of the mass fractions of \nuc{Si}{28}, 
    \nuc{S}{32}, \nuc{Ar}{36}, \nuc{Ca}{40}, and \nuc{Ti}{44}.  The Iron group 
    mass fraction is the sum of the mass fractions of \nuc{Cr}{48}, 
    \nuc{Fe}{52}, \nuc{Ni}{56}, and \nuc{Zn}{60}}
	\label{fig:xc67}
\end{figure}

\begin{table}
    \centering 
    \caption{Comparison of network abundances at feezeout with NSE for \tnine=6 
    and $\rho=10^7 \gcc$}
    \label{tab:nse}
    \vspace{12pt}
    \begin{tabular}{c|lll}
        \tableline
        Nucleus & \multicolumn{1}{c}{$Y_{net}$}&
        \multicolumn{1}{c}{$Y_{qse}$}& \multicolumn{1}{c}{$Y_{nse}$} \\
        \tableline
        \nuc{He}{4} & 1.23\ttt{-1}& 1.24\ttt{-1}& 1.25\ttt{-1}\\
        \nuc{C}{12} & 1.03\ttt{-6}& 1.04\ttt{-6}& 1.04\ttt{-6}\\
        \nuc{O}{16} & 1.88\ttt{-6}& 1.91\ttt{-6}& 1.89\ttt{-6}\\
        \nuc{Ne}{20}& 3.54\ttt{-8}& 3.62\ttt{-8}& 3.56\ttt{-8}\\
        \nuc{Mg}{24}& 4.08\ttt{-6}& 4.18\ttt{-6}& 4.08\ttt{-6}\\
        \nuc{Si}{28}& 1.28\ttt{-3}& 1.32\ttt{-3}& 1.28\ttt{-3}\\
        \nuc{S}{32} & 1.21\ttt{-3}& 1.23\ttt{-3}& 1.20\ttt{-3}\\
        \nuc{Ar}{36}& 7.03\ttt{-4}& 7.13\ttt{-4}& 7.00\ttt{-4}\\
        \nuc{Ca}{40}& 7.60\ttt{-4}& 7.64\ttt{-4}& 7.55\ttt{-4}\\
        \nuc{Ti}{44}& 4.16\ttt{-5}& 4.14\ttt{-5}& 4.12\ttt{-5}\\
        \nuc{Cr}{48}& 2.70\ttt{-4}& 2.71\ttt{-4}& 2.67\ttt{-4}\\
        \nuc{Fe}{52}& 1.43\ttt{-3}& 1.42\ttt{-3}& 1.41\ttt{-3}\\
        \nuc{Ni}{56}& 5.13\ttt{-3}& 5.06\ttt{-3}& 5.06\ttt{-3}\\
        \nuc{Zn}{60}& 2.98\ttt{-6}& 2.92\ttt{-6}& 2.94\ttt{-6}\\
        \tableline
    \end{tabular}
\end{table}

For higher temperatures, large abundances of free \alp-particles reduce the 
importance of the silicon and iron peak nuclei.  As Column 4 of 
Table~\ref{tab:nse} reveals, even the NSE abundance distribution which 
represents the end point of silicon burning may be dominated by free 
\alp-particles.  In spite of the larger role played by free \alp-particles, 
the QSE-reduced \alp\ network successfully replicates the nuclear evolution 
during silicon burning, because the nuclei of the silicon and iron peak 
groups still obey QSE. Fig.~\ref{fig:xc67} compares the nuclear evolution 
of the independant mass fractions from the QSE-reduced \alp\ network to 
their conventional counterparts, with \tnine=6 and $\rho=10^{7} \gcc$.  
Here too, the iron group mass fraction is initially over-predicted by the 
QSE-reduced network, though to a lesser extent (7\% after $10^{-8}$ 
seconds).  This over-prediction also subsides as the iron peak abundances 
increase, and hence their adherence to QSE improves.  As in the previous 
case, despite these small differences in abundance, the agreement between 
the rates of energy generation is better than 1\%.  Examination of 
Table~\ref{tab:nse} reveals that the equilibrium abundances for the 
individual nuclei calculated by the QSE-reduced network also agree well 
with those calculated by the conventional network and by NSE. Even for the 
smallest abundances, the differences are less than 3\%.

While examples of silicon burning calculations under constant conditions are
instructive, if the QSE-reduced \alp\ network is to replace the conventional 
\alp\ network it must be shown to be accurate under changing
thermodynamic conditions.  Of particular importance is the ability to
model explosive silicon burning.  As a representative 
analytic model for silicon burning occurring as a result of shock heating, 
we will consider a mass zone which is instantaneously heated by a passing 
shock to some peak initial temperature \tninei\ and density \rhoi\, and 
then expands and cools adiabatically.  Following the approximation 
introduced by \cite{FoHo64}, the hydrodynamic expansion timescale (equal 
to the free fall timescale) is
\begin{equation}
\tau_{\rm HD} = \left(24 \pi G \rho \right)^{-1/2} = 446 \rho_6^{-1/2} 
{\rm ms}, 
\end{equation}
with the time dependence of the density and temperature of this
radiation dominated gas given by
\begin{eqnarray}
\rho(t) &= &\rho_i\exp \left( -t / \tau_{\mathrm HD} \right), \nonumber \\
T_9(t) &= &T_{9i} \exp \left( -t / 3 \tau_{\mathrm HD} \right), 
\end{eqnarray}
assuming the adiabatic exponent $\gamma= 4/3$.  For initial conditions
we chose conditions identical to the examples discussed above.

\begin{figure}[tbp]
    \centering
    \includegraphics[angle=90,width=\textwidth]{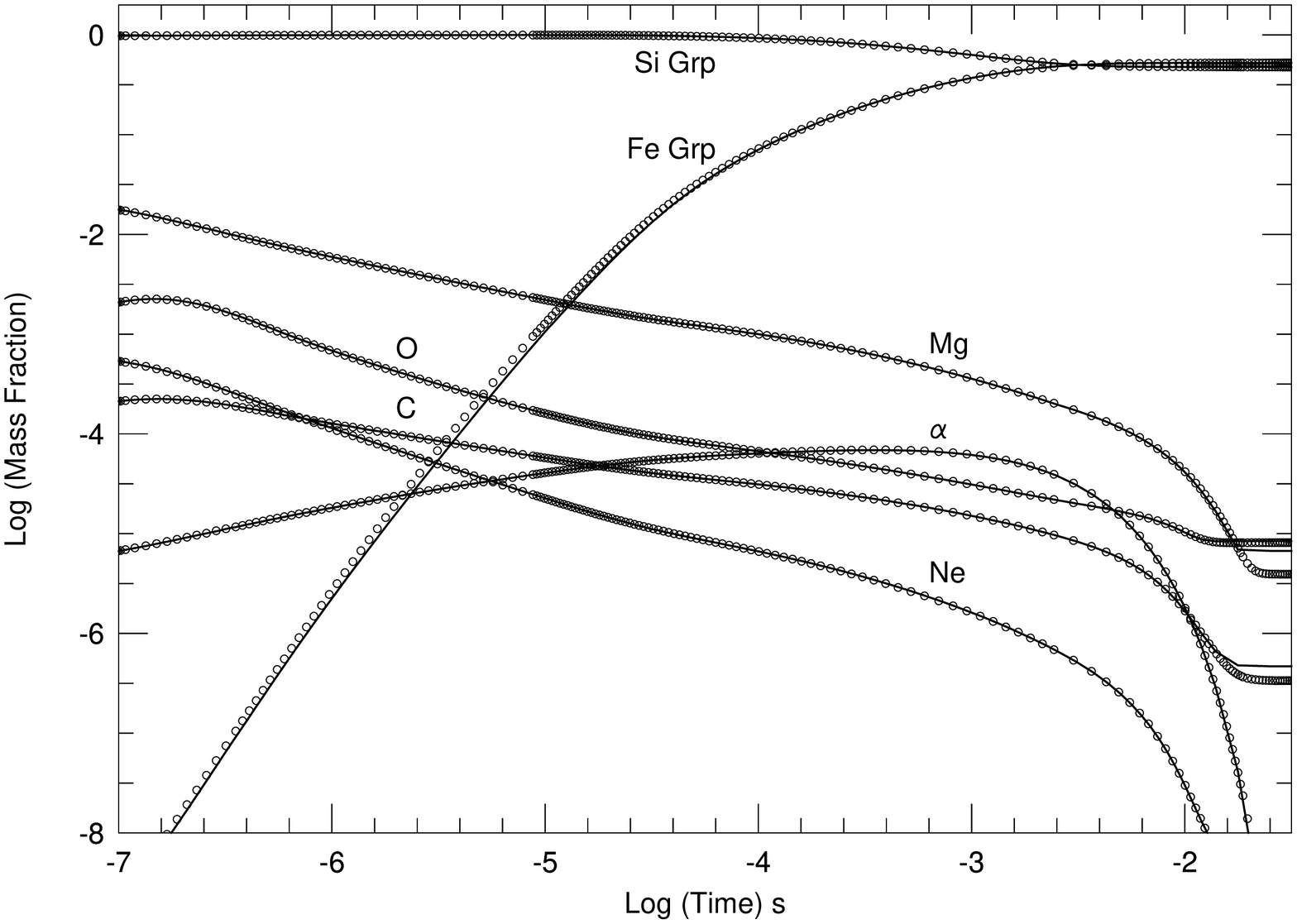} 
    \caption{Evolution of the independant nuclear mass fractions under 
    adiabatic expansion with \tninei=5 and $\rhoi= 10^{9} \gcc$.  The 
    solid lines display the evolution due to a conventional \alp\ 
    network, the circles show the evolution by the QSE-reduced \alp\ 
    network.  The Silicon group mass fraction is the sum of the mass 
    fractions of \nuc{Si}{28}, \nuc{S}{32}, \nuc{Ar}{36}, \nuc{Ca}{40}, 
    and \nuc{Ti}{44}.  The Iron group mass fraction is the sum of the 
    mass fractions of \nuc{Cr}{48}, \nuc{Fe}{52}, \nuc{Ni}{56}, and 
    \nuc{Zn}{60}}
    \label{fig:xe59} 
\end{figure}

Figure~\ref{fig:xe59} shows the nuclear evolution for an example of this 
explosive burning model with $\tninei=5$ and $\rhoi=10^9 \gcc$.  Over the 
first millisecond, the evolution portrayed here is virtually identical to 
that of Figure~\ref{fig:xc59}; however by the time one millisecond has 
elapsed, the temperature has dropped to 4.9 \ttt{9} \kel, slowing the 
reactions which are turning silicon into iron peak elements.  This cooling, 
which drops \tnine\ below 4 after 9 ms and below 3 after 22 ms, freezes out 
the nuclear reactions before NSE is reached, resulting in incomplete 
silicon burning, as discussed by WAC. In this case, the freezeout leaves 
nearly equal amounts of silicon group and iron peak group elements.  
Throughout most of the evolution in this example, the agreement between the 
mass fractions as evolved by the QSE-reduced \alp\ network with those 
evolved by its conventional counterpart is comparable to that demonstrated 
under constant thermodynamic conditions.  Columns 2 and 3 of 
Table~\ref{tab:nfreeze} compare the abundances after 9 ms have elapsed, 
with \tnine\ nearing 4.  In this case, as for that displayed in 
Table~\ref{tab:const1}, the largest relative error (5\%) is in the 
abundance of \nuc{Cr}{48}.  These small differences in abundance result in 
small differences in the accumulated nuclear energy release, approximately 
.5\% to this point.  By this time, adiabatic cooling has greatly reduced 
the rate of energy generation from its peak of more than $10^{22} \erggs$ 
to roughly $10^{17} \erggs$.  Though the absolute difference in the rate of 
energy generation as calculated by the 2 networks has declined from $\sim 
10^{19}$ to $10^{16} \erggs$, the relative difference has grown to 10\% as 
\tnine\ nears 4.  Fortunately this difference is negligible, since nuclear 
energy release has virtually ceased, with $<.2\%$ of the total energy 
release remaining.

\begin{table}
    \centering 
    \caption{Comparison of network abundances for \tninei=5.0 and \rhoi= $10^9 \gcc$}
    \label{tab:nfreeze}
    \vspace{12pt}
    \begin{tabular}{c|lllll}
        \tableline
        Time (ms)&\multicolumn{2}{c}{8.77}&
        \multicolumn{2}{c}{17.7}&\multicolumn{1}{c}{255} \\
        \tnine& \multicolumn{2}{c}{4.07}&
        \multicolumn{2}{c}{3.29}&\multicolumn{1}{c}{0.01} \\
        \tableline \tableline
        Nucleus & \multicolumn{1}{c}{$Y_{net}$}
        &\multicolumn{1}{c}{$Y_{qse}$}& \multicolumn{1}{c}{$Y_{net}$} 
        &\multicolumn{1}{c}{$Y_{qse}$}& \multicolumn{1}{c}{$Y_{net}$} \\
        \tableline
\nuc{He}{4} & 7.90\ttt{-7} & 7.82\ttt{-7} & 1.04\ttt{-8} & 1.01\ttt{-8} & 1.94\ttt{-14}\\
\nuc{C}{12} & 1.96\ttt{-7} & 1.96\ttt{-7} & 3.99\ttt{-8} & 3.23\ttt{-8} & 3.90\ttt{-8} \\
\nuc{O}{16} & 7.34\ttt{-7} & 7.39\ttt{-7} & 5.26\ttt{-7} & 5.07\ttt{-7} & 5.27\ttt{-7} \\
\nuc{Ne}{20}& 2.63\ttt{-9} & 2.63\ttt{-9} & 1.69\ttt{-10}& 1.48\ttt{-10}& 9.88\ttt{-11}\\
\nuc{Mg}{24}& 2.24\ttt{-6} & 2.26\ttt{-6} & 2.88\ttt{-7} & 2.98\ttt{-7} & 2.80\ttt{-7} \\
\nuc{Si}{28}& 7.65\ttt{-3} & 7.76\ttt{-3} & 7.58\ttt{-3} & 7.86\ttt{-3} & 7.58\ttt{-3} \\
\nuc{S}{32} & 4.93\ttt{-3} & 4.96\ttt{-3} & 5.16\ttt{-3} & 5.15\ttt{-3} & 5.16\ttt{-3} \\
\nuc{Ar}{36}& 1.43\ttt{-3} & 1.42\ttt{-3} & 1.27\ttt{-3} & 1.21\ttt{-3} & 1.27\ttt{-3} \\
\nuc{Ca}{40}& 1.32\ttt{-3} & 1.30\ttt{-3} & 1.32\ttt{-3} & 1.22\ttt{-3} & 1.32\ttt{-3} \\
\nuc{Ti}{44}& 7.07\ttt{-6} & 6.90\ttt{-6} & 1.96\ttt{-6} & 1.72\ttt{-6} & 1.69\ttt{-6} \\
\nuc{Cr}{48}& 5.89\ttt{-5} & 5.58\ttt{-5} & 4.40\ttt{-5} & 1.19\ttt{-5} & 4.40\ttt{-5} \\
\nuc{Fe}{52}& 7.17\ttt{-4} & 6.93\ttt{-4} & 6.33\ttt{-4} & 3.05\ttt{-4} & 6.33\ttt{-4} \\
\nuc{Ni}{56}& 8.63\ttt{-3} & 8.60\ttt{-3} & 8.73\ttt{-3} & 9.04\ttt{-3} & 8.73\ttt{-3} \\
\nuc{Zn}{60}& 6.18\ttt{-8} & 6.06\ttt{-8} & 3.26\ttt{-9} & 3.23\ttt{-9} & 4.38\ttt{-10}\\
        \tableline
    \end{tabular}
\end{table}

Though energetically unimportant, nuclear reactions continue, resulting in 
significant changes in the smaller abundances.  As pointed out by WAC, and 
discussed by HT98 in the context of a large nuclear network, the continued 
cooling results in the gradual freezeout of the photodisintegrations 
responsible for QSE.  Typically, a photodisintegration only represents a 
significant flux if its Q value is less than $30 \thinspace k_B T \sim 2.6 
\thinspace \tnine\ \mev$.  The \alp\ captures which link silicon to nickel 
typically have Q values of 7-8 \mev, thus QSE begins to fail as \tnine\ 
approaches 3.  However this same decline in temperature also reduces the 
rate of charged particle capture reactions, greatly reducing the amount of 
nucleosynthesis which occurs after \tnine\ drops below $\sim3.5$.  The 
group abundances of the silicon and iron peak groups (which account for 
99.9\% of the mass), as calculated by the QSE-reduced \alp\ network after 
18 ms have elapsed (\tnine=3.3), differ by less than 1\% from those of the 
conventional \alp\ network at the same point in time.  Comparison of 
Columns 4 and 5 of Table~\ref{tab:nfreeze} reveals larger variations among 
individual abundances, most notably, significant under estimation by the 
QSE-reduced network of the \nuc{Cr}{48} and \nuc{Fe}{52} abundances, with a 
small, compensatory over estimate of the \nuc{Ni}{56} abundance.  These 
variations, factors of 2 and 4 for \nuc{Fe}{52} and \nuc{Cr}{48}, 
respectively, and 3\% for \nuc{Ni}{56} signal the breakdown of the iron 
peak QSE group.  With the steep decline in temperature and density, the 
flux upward from \nuc{Fe}{52} in the conventional network is no longer 
sufficient to provide the reduction in abundance which QSE and the sharply 
declining abundance of free \alp-particles requires.

As the temperature and density continue to decline, so to does the free
\alp-particle abundance.  Column 6 of Table~\ref{tab:nfreeze} details the 
abundances after 255 ms have elapsed, with \tnine\ having dropped to .01, 
and all abundances having frozen out.  Comparison of colums 4 and 6
reveals that the decline of the free \alp-particle abundance by $10^{-8}$ 
is the largest abundance variation beyond 18 ms.  Since the more abundant
species have effectively frozen out by the time \tnine\ approaches 3.5,
comparison of columns 5 and 6 reveal that the predictions of the QSE-reduced 
network, frozen near \tnine=3.5, also provide good abundance estimates, 
in addition to the excellent estimates of the rate of energy generation
discusssed earlier, in spite of the freezeout.  

\begin{figure}[tbp]
	\centering
	\includegraphics[angle=90,width=\textwidth]{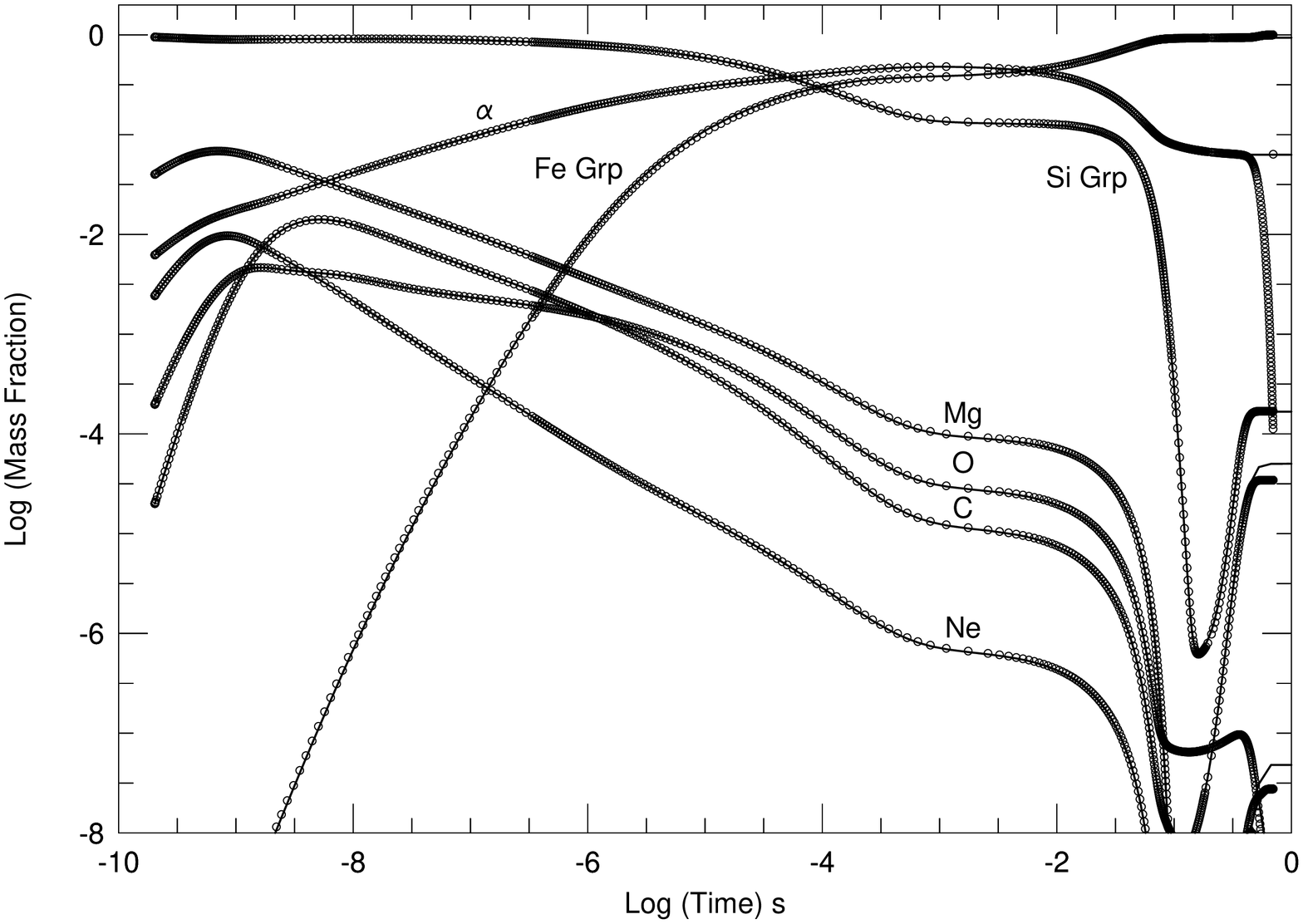} 
	\caption{Evolution of the independant nuclear mass fractions under 
	adiabatic expansion with \tninei=6 and $\rhoi=10^{7} \gcc$.  The solid 
	lines display the evolution due to a conventional \alp\ network, the 
	circles show the evolution by the QSE-reduced \alp\ network.  The 
	Silicon group mass fraction is the sum of the mass fractions of 
	\nuc{Si}{28}, \nuc{S}{32}, \nuc{Ar}{36}, \nuc{Ca}{40}, and \nuc{Ti}{44}.
	The Iron group mass fraction is the sum of the mass fractions of
	\nuc{Cr}{48}, \nuc{Fe}{52}, \nuc{Ni}{56}, and \nuc{Zn}{60}}
	\label{fig:xe67} 
\end{figure}

Figure~\ref{fig:xe67} shows the evolution of another example of this 
explosive silicon burning model with $\tninei=6$ and $\rhoi=10^7 \gcc$.  
The lower density in this case results in a slower expansion, with \tnine\ 
reaching 5, 4, and 3 after 77, 172, and 293 milliseconds, respectively.  
The nuclear evolution is similar to that portrayed in Figure~\ref{fig:xc67} 
for the first 10 milliseconds, after which the declining temperature favors 
the iron peak group at the expense of lighter nuclei.  Though the rate of 
cooling is relatively slow, the temperature drops too quickly for the large 
\alp-particle abundance to be completely incorporated into heavier nuclei, 
resulting in \emph{\alp-rich freezeout} (see \cite{ArTW71}, and 
\cite{WoAC73}).  Even with this large overabundance of \alp-particles, the 
QSE-reduced network tracks the group abundances and rate of energy 
generation as reliably as it did under constant conditions until \tnine\ 
approaches 3.5.  By the time 223 ms have elapsed (\tnine=3.5), the largest 
relative difference, occuring in the small ($\sim 10^{-6}$) abundance of 
the silicon group, has grown to 9\%.  With only such small errors in 
abundance, it is not surprising that the difference in the rate of energy 
generation remains less than 1\%.  There is greater disparity between the 
individual abundances as calculated by the QSE-reduced network and those 
calculated by the conventional network.

\begin{table}
    \centering 
    \caption{Comparison of network abundances near freezeout for \tninei=6.0 
    and \rhoi=$10^7 \gcc$}
    \label{tab:afreeze}
    \vspace{12pt}
    \begin{tabular}{c|lllll}
        \tableline
        Time (s)&\multicolumn{2}{c}{.174}&
        \multicolumn{2}{c}{.290}&\multicolumn{1}{c}{2.67} \\
        \tnine& \multicolumn{2}{c}{3.98}&
        \multicolumn{2}{c}{3.02}&\multicolumn{1}{c}{0.01} \\
        \tableline \tableline
        Nucleus & \multicolumn{1}{c}{$Y_{net}$}
        &\multicolumn{1}{c}{$Y_{qse}$}& \multicolumn{1}{c}{$Y_{net}$}
        & \multicolumn{1}{c}{$Y_{qse}$}& \multicolumn{1}{c}{$Y_{net}$} \\
        \tableline
\nuc{He}{4} & 1.73\ttt{-2} & 1.71\ttt{-2} & 1.62\ttt{-2} & 1.61\ttt{-2} & 1.57\ttt{-2} \\
\nuc{C}{12} & 1.60\ttt{-9} & 1.58\ttt{-9} & 9.76\ttt{-8} & 9.62\ttt{-8} & 4.18\ttt{-6} \\
\nuc{O}{16} & 4.24\ttt{-9} & 4.19\ttt{-9} & 5.51\ttt{-9} & 5.45\ttt{-9} & 2.16\ttt{-10}\\
\nuc{Ne}{20}& 8.55\ttt{-12}& 8.46\ttt{-12}& 7.05\ttt{-11}& 6.96\ttt{-11}& 2.92\ttt{-10}\\
\nuc{Mg}{24}& 4.38\ttt{-12}& 4.33\ttt{-12}& 3.25\ttt{-11}& 3.21\ttt{-11}& 2.00\ttt{-9} \\
\nuc{Si}{28}& 1.92\ttt{-11}& 6.39\ttt{-14}& 1.92\ttt{-10}& 8.10\ttt{-22}& 2.95\ttt{-9} \\
\nuc{S}{32} & 4.94\ttt{-11}& 4.42\ttt{-12}& 5.34\ttt{-10}& 2.14\ttt{-17}& 2.08\ttt{-8} \\
\nuc{Ar}{36}& 2.60\ttt{-10}& 1.26\ttt{-10}& 2.46\ttt{-9} & 1.73\ttt{-13}& 1.59\ttt{-7} \\
\nuc{Ca}{40}& 1.19\ttt{-8} & 1.12\ttt{-8} & 1.90\ttt{-8} & 6.44\ttt{-9} & 4.18\ttt{-7} \\
\nuc{Ti}{44}& 4.73\ttt{-9} & 4.68\ttt{-9} & 1.71\ttt{-7} & 1.69\ttt{-7} & 3.26\ttt{-6} \\
\nuc{Cr}{48}& 1.88\ttt{-8} & 1.51\ttt{-8} & 2.19\ttt{-8} & 1.26\ttt{-14}& 2.55\ttt{-6} \\
\nuc{Fe}{52}& 1.59\ttt{-5} & 1.61\ttt{-5} & 9.26\ttt{-7} & 1.35\ttt{-8} & 4.50\ttt{-6} \\
\nuc{Ni}{56}& 1.66\ttt{-2} & 1.66\ttt{-2} & 1.67\ttt{-2} & 1.67\ttt{-2} & 1.64\ttt{-2} \\
\nuc{Zn}{60}& 6.50\ttt{-6} & 6.39\ttt{-6} & 4.23\ttt{-5} & 4.15\ttt{-5} & 2.97\ttt{-4} \\
        \tableline
    \end{tabular}
\end{table}

As discussed by HT98 for a more complete nuclear network, in cases of 
\alp-rich freezeout the large \alp-particle abundance results in a 
relatively large nuclear flow upward into the silicon group.  This flow 
gradually disrupts the QSE groups by increasing the abundance of the less 
massive members of the group while their QSE abundances drop sharply.  
Fortunately for our use of QSE to reduce the network, because of the large 
free \alp-particle abundance, these less massive members have only small 
abundances and thus the effects of this disruption are small.  Comparison 
of abundances calculated by the conventional \alp\ network after 170 ms 
have elapsed $(\tnine \sim 4)$ with those of its QSE-reduced counterpart 
(columns 2 and 3 of Table~\ref{tab:afreeze}) shows the beginning of this 
process, with the abundances of \nuc{Si}{28} and \nuc{S}{32} much larger 
than QSE would predict.  In spite of this breakdown in QSE, the abundances 
predicted for the more abundant members of the group and, as a result, the 
rate of energy generation, agree well.  However as the temperature 
continues to drop, the disruption of the QSE groups by the large free 
\alp-particle abundance grows.  By the time \tnine\ drops to 3 (columns 4 
\& 5 of Table~\ref{tab:afreeze}), the under prediction by QSE of the 
smaller group abundances also effects the iron peak group.  The abundances 
of the dominant nuclei, \alp, \nuc{Ni}{56} and \nuc{Zn}{60}, and hence the 
energy production, are still in good agreement, with the energy generation 
predictions differing by less than 1\%.

Of particular importance, the abundance of \nuc{Zn}{60} is within 2\% of 
that predicted by QSE. With a Q value of $2.7 \mev$, $\nuc{Ni}{56}+ \alp 
\leftrightarrow \nuc{Zn}{60} + \gamma$ remains balanced until significantly 
later than the reaction pairs connecting silicon to nickel.  As the 
remaining photodisintegrations freeze out, the continued capture of the 
large abundance of \alp-particles results in a large number of \alp\ 
captures and significant abundances changes, the most prominent of which is 
the conversion of \nuc{Ni}{56} to \nuc{Zn}{60}.  As late as an elapsed time 
of .35 seconds, with \tnine=2.6 and $Y(\nuc{Zn}{60})=1.5 \ttt{-4}$, the 
abundance calculated for \nuc{Zn}{60} by the QSE-reduced network is within 
5\% of that predicted by the conventional network.  As a result of this 
continued validity of QSE for the abundance of \nuc{Zn}{60}, the energy 
generation rate predicted by the QSE-reduced \alp\ network at an elapsed 
time of .35 seconds is only 11\% greater than that calculated by the full 
\alp\ network.  As the temperature continues to drop, the continued \alp\ 
captures disrupt this last balanced reaction pair, converting even more 
\nuc{Ni}{56} into \nuc{Zn}{60}.  Column 6 of Table~\ref{tab:afreeze} shows 
the abundances after 2.7 seconds have elapsed, by which time all of the 
abundances are frozen.  While virtually all of the abundances have grown at 
the expense of those of free \alp-particles and \nuc{Ni}{56}, the most 
significant is a doubling of the abundance of \nuc{Zn}{60} after 
.35 seconds.  Though the QSE-reduced network can not track the nuclear 
evolution to this point, the relatively small amount of energy released, 
less than 1\% of that released over the first .35 seconds, makes this only 
a minor weakness.

\section{The \alp7 network}

While the preceding sections demonstrate the success of the QSE-reduced 
\alp\ network as a replacement for a conventional 
\alp\ network during silicon burning, it can not accurately calculate the 
evolution of \yf\ when QSE does not apply.  Prior to silicon burning, the 
primary abundances of interest are those of \alp, \nuc{C}{12}, \nuc{O}{16}, 
\nuc{Ne}{20}, \nuc{Mg}{24}, and \nuc{Si}{28}.  This subset of nuclei, which we 
will call \calRp, is identical to \calR, less \nuc{Ni}{56}.  Thus evolution 
of a 
conventional \alp\ network for the 7 nuclei of the set \calR\ can model these 
earlier burning stages nearly as well as the larger \alp\ network which evolves 
the entire set \calF, with the only difficulty being small abundances of nuclei 
heavier than silicon.  In combination with the QSE-reduced \alp\ network, one 
could hope to globally replace the full \alp\ network with a smaller network 
which evolves \yr, using QSE where it applies.  The principle uncertainty in 
the use of this combined network, which we have dubbed \alp 7, is the the 
nature and timing of the transition to QSE.

The first necessary condition for QSE is that the temperature be in excess 
of 3 \ttt{9} \kel, because, as discussed in \S 3, for lower temperatures 
the rates of photodisintegration are much smaller than the corresponding 
captures.  An additional requirement for QSE to provide a good estimate of 
the abundances of heavier elements is that a significant fraction of the 
matter be composed of nuclei with $A>24$.  As long as the material is 
principally composed of lighter nuclei, like \nuc{C}{12}, \nuc{O}{16}, or 
\nuc{Ne}{20}, the production and destruction of free \alp-particles is 
dominated by reactions among these nuclei, rendering suspect QSE abundance 
predictions for the silicon and iron peak group nuclei based on these free 
\alp-particle abundances.  Once these conditions are met, the abundances of 
the set \calRp\ must be mapped in a suitable way into the initial 
abundances for the set \calG\ which the QSE-reduced \alp\ network evolves.

\begin{figure}[tbp]
	\centering
	\includegraphics[angle=90,width=\textwidth]{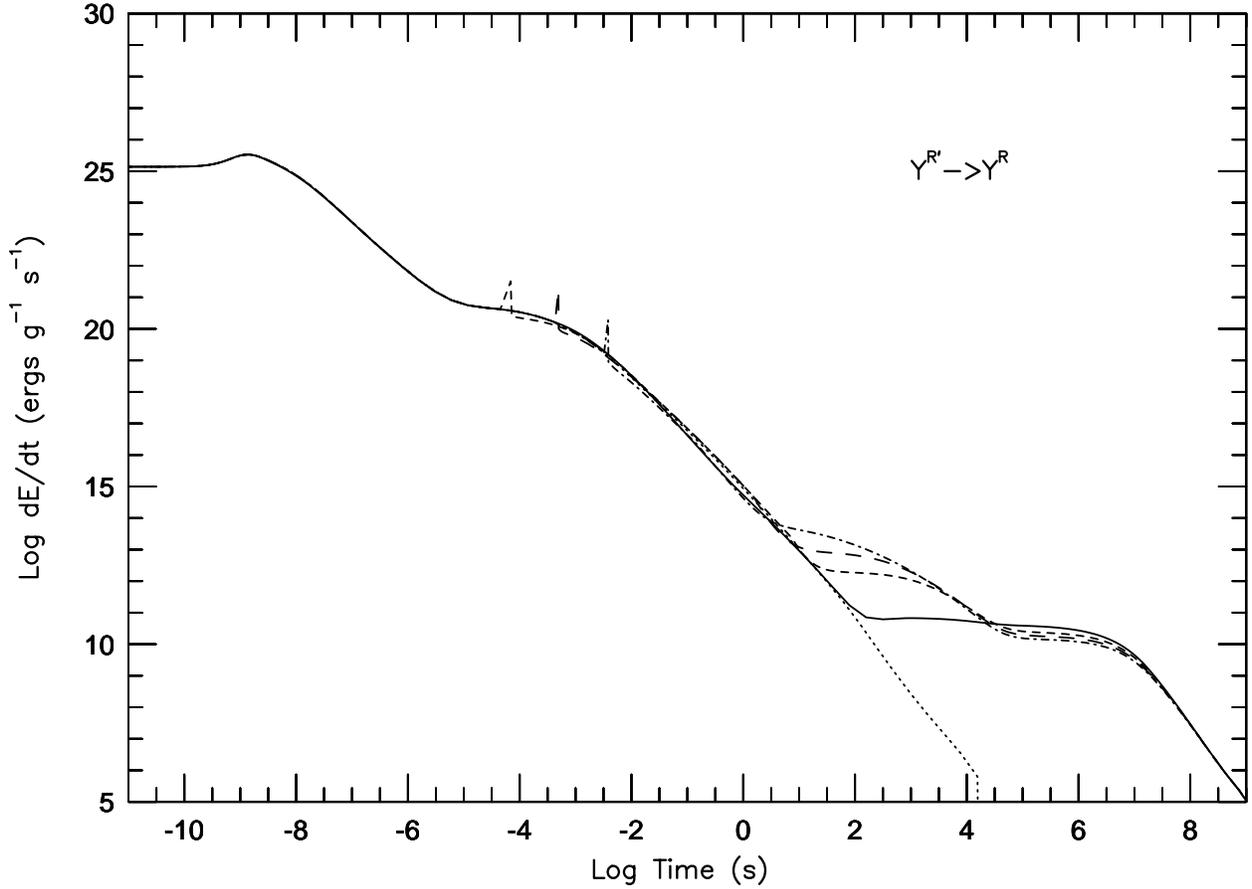} 
	\caption{Comparison of the Energy generation by the full \alp\ network 
	under constant thermodyamic conditions \tnine=3 and $\rho=10^{9} \gcc$ 
	to the \alp 7 network, assuming the individual abundance transition.  
	The solid curve is the result of the conventional 14 element \alp\ 
	network. The dotted curve shows the results of the \alp 7 network, with 
	the transtion to QSE turned off.  The short dashed, long dashed and dot 
	dashed curves display the results of the \alp 7 network, with the 
	transition to QSE occuring when approximately 30\%, 60\% and 90\% of
	the mass is in nuclei heavier than $A=24$, respectively.} 
	\label{fig:eqq} 
\end{figure}

\begin{figure}[tbp]
	\centering
	\includegraphics[angle=90,width=\textwidth]{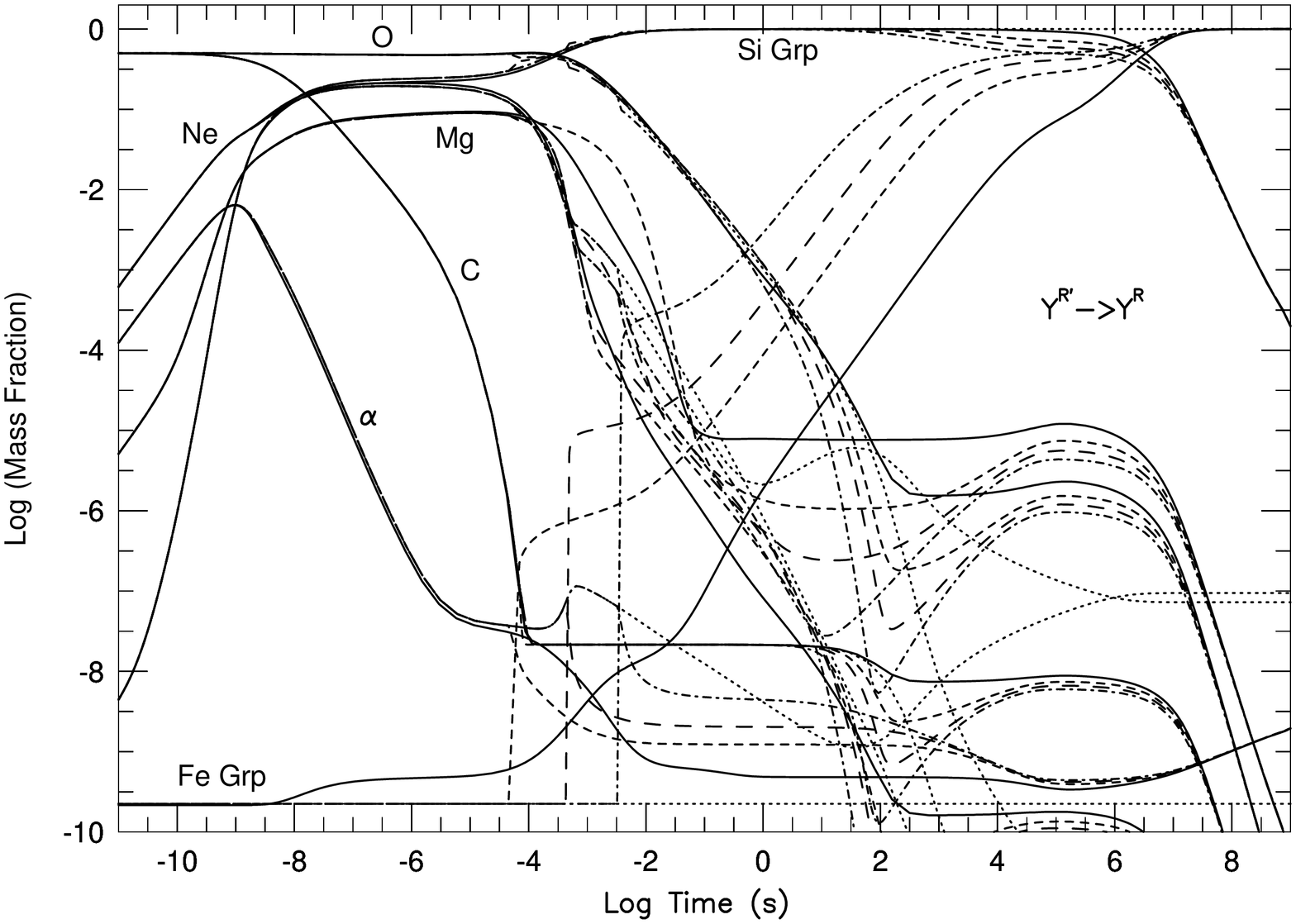} 
	\caption{Comparison of the evolution of the independant mass fractions 
	by the full \alp\ network under constant thermodyamic conditions 
	\tnine=3 and $\rho= 10^{9} \gcc$ to the \alp 7 network, assuming the 
	individual abundance transition.  The solid curves are the result of 
	the conventional 14 element \alp\ network. The dotted curves show the 
	results of the \alp 7 network with the transtion to QSE turned off.  
	The short dashed, long dashed and dot dashed curves display the results 
	of the \alp 7 network, with the transition to QSE occuring when 
	approximately 30\%, 60\% and 90\% of the mass is in nuclei heavier 
	than $A=24$, respectively. } 
	\label{fig:xqq} 
\end{figure}

One approach to this transition seeks to conserve the individual abundances 
of the sets \calR\ and \calRp\ across the transition, setting $\yrp = \yr$, 
and then calculating the initial values for \yg\ according to Eq.~\ref{eq:yg}. 
Figure~\ref{fig:eqq} displays the energy generation under constant conditions 
(\tnine=3 and $\rho=10^9 \gcc$) for the \alp 7 network using this \emph{
individual abundance} (IA) approach to the transition.  Figure~\ref{fig:xqq} 
shows the group abundances for this same IA case.  For Figures 5-8, the 
solid curves are the result of the conventional 14 element \alp\ network 
and the dotted curves show the results of the 7 element \alp\ network alone. 
The short dashed, long dashed and dot dashed curves display the results of 
the combined \alp 7 network, with the transition to QSE occuring when 
approximately 30\%, 60\% and 90\% of the mass is in nuclei heavier than 
$A=24$, respectively.  The spikes in the short dashed, long dashed and dot 
dashed curves seen in Fig.~\ref{fig:eqq} result from errors in mass 
conservation at the transition.  Treating \yrp\ as \yr\ results in an over 
estimate of the silicon and iron peak group mass fractions at the expense 
of \nuc{O}{16}, producing the spike in energy generation.  This overabundance 
of iron also causes the excess energy production around 100 s and the 
corresponding dearth around $10^6$ s.

\begin{figure}[tbp]
	\centering
	\includegraphics[angle=90,width=\textwidth]{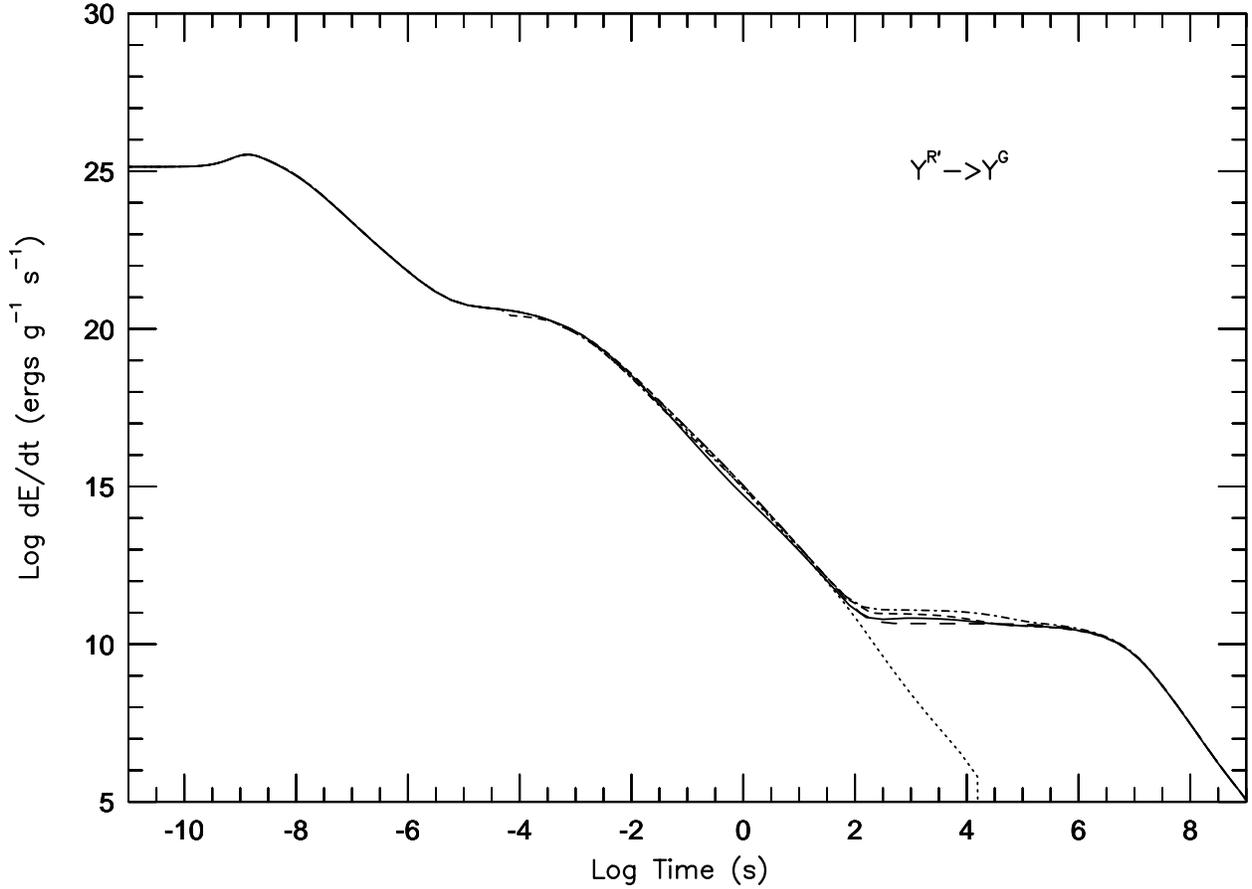} 
	\caption{Comparison of the Energy generation by the full \alp\ network 
	under constant thermodyamic conditions \tnine=3 and $\rho=10^{9} \gcc$ 
	to the \alp 7 network, assuming the group abundance transition.  The 
	solid curve is the result of the conventional 14 element \alp\ network. 
	The dotted curve shows the results of the \alp 7 network, with the 
	transtion to QSE turned off.  The short dashed, long dashed and dot 
	dashed curves display the results of the \alp 7 network, with the 
	transition to QSE occuring when approximately 30\%, 60\% and 90\% of
	the mass is in nuclei heavier than $A=24$, respectively. } 
	\label{fig:eqr} 
\end{figure}

\begin{figure}[tbp]
	\centering
	\includegraphics[angle=90,width=\textwidth]{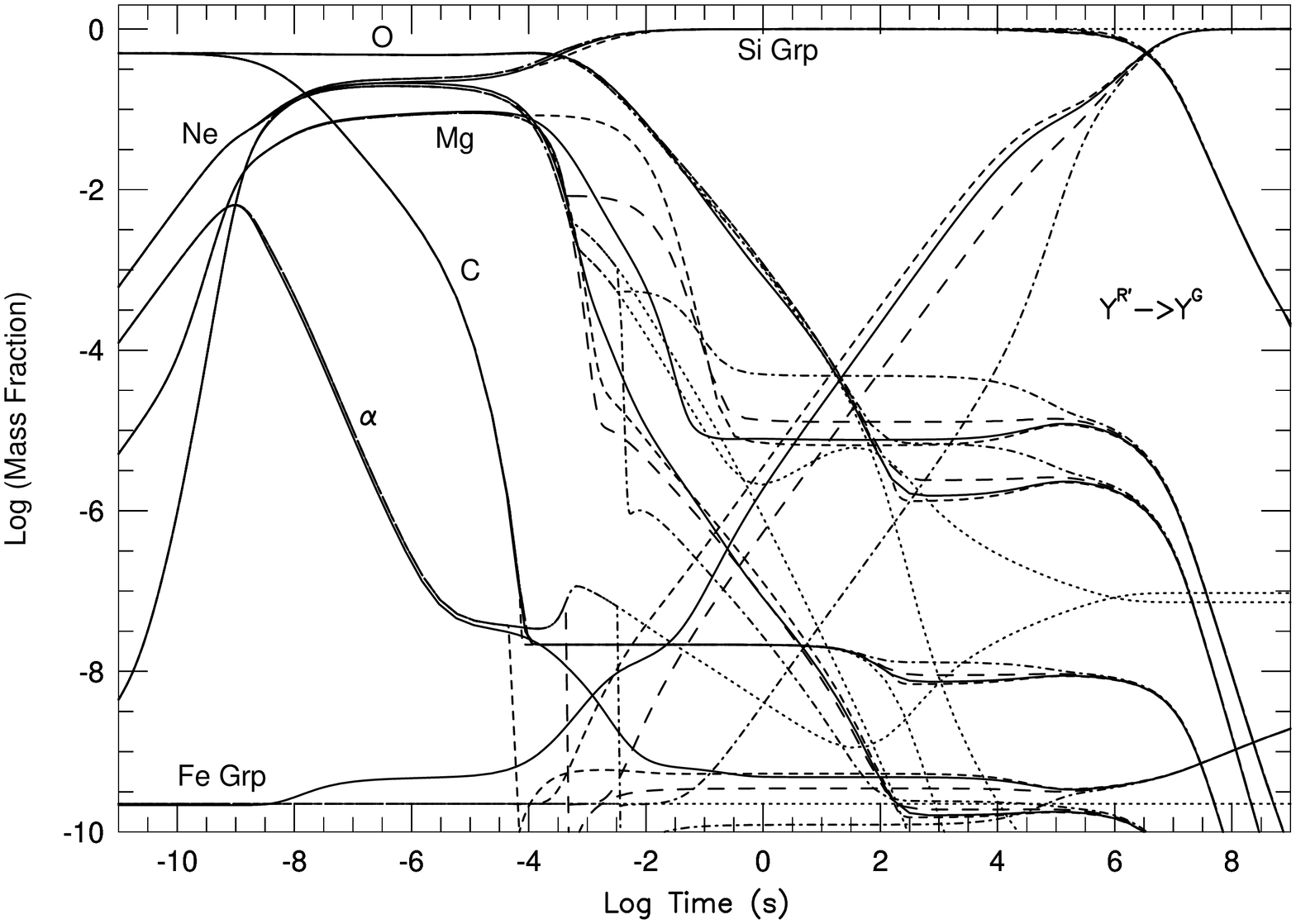} 
	\caption{Comparison of the evolution of the independant mass fractions 
	by the full \alp\ network under constant thermodyamic conditions 
	\tnine=3 and $\rho= 10^{9} \gcc$ to the \alp 7 network, assuming the 
	group abundance transition.  The solid curves are the result of the 
	conventional 14 element \alp\ network. The dotted curves show the results 
	of the \alp 7 network with the transtion to QSE turned off.  The short 
	dashed, long dashed and dot dashed curves display the results of the 
	\alp 7 network, with the transition to QSE occuring when approximately 
	30\%, 60\% and 90\% of the mass is in nuclei heavier than $A=24$, 
	respectively. } 
	\label{fig:xqr} 
\end{figure}

If one examines the \alp-particle abundance (shown in Fig.~\ref{fig:xqq}) 
in the vicinity of $10^{-3}$ s, one sees that the 7 element \alp\ network, 
and its \alp 7 network descendants, begin to overproduce free 
\alp-particles.  In the conventional 14 element \alp\ network, these 
\alp-particles are capturing on \nuc{Si}{28}, populating the Si group at 
the expense of \nuc{Si}{28}, an option not allowed the smaller network.  It 
is these overlarge abundaces of \nuc{Si}{28} and free \alp-particles which 
cause the errors in mass conservation when QSE is first applied.  A 
successful transition requires awareness that the abundance of \nuc{Si}{28} 
in the set \calRp\ represents the total of all abundances above 
\nuc{Mg}{24}.  In terms of \yg, the capture of an \alp-particle on 
\nuc{Si}{28} does not change $Y_{\alpha G}$ or $Y_{Si G}$.  If we consider 
\yg\ when nuclei heavier than \nuc{Si}{28} are not evolved, the sums in 
Eq.~\ref{eq:yg} disappear, leaving $ Y_{\alpha G} = Y_{\alpha}$, $Y_{Si G} 
= Y(\nuc{Si}{28})$, and $Y_{Fe G} = Y(\nuc{Ni} {56})$ = constant.  Thus 
\yrp\ can also be mapped directly to \yg.  Since \yrp\ and \yg\ have the 
same normalization rule, mass is conserved.  This approach seeks to 
conserve the abundances of the groups across the QSE transition, hence we 
refer to this as the \emph{group abundance} (GA) transition.  
Figure~\ref{fig:eqr} shows the energy generation, again for burning at a 
constant \tnine=3 and $\rho=10^9 \gcc$, using this approach to transition.  
The results agree more closely with the conventional \alp\ network for all 
three values of the transition mass fraction, with the best results 
occuring for a transition mass fraction around 50\%.  The mass fractions 
predicted by this approach, shown in Figure~\ref{fig:xqr}, are also more 
reliable, though there are still significant discrepancies among the 
smaller mass fractions, particularly near the transition to QSE and for a 
transition mass fraction of 90\%.

\section{Conclusion}

In this paper, we have demonstrated the ability of QSE to reduce the number 
of independant nuclear abundances which need to be evolved to track 
\alp-chain nucleosynthesis for silicon burning.  An \alp\ network reduced 
in this fashion is capable of providing reasonable estimates of the nuclear 
energy generation and elemental production, even in cases where 
thermodynamic variations result in incomplete silicon burning or a 
freezeout rich in free \alp-particles.  Computational, this reduction in 
the number of independent nuclear abundance variables offers two 
advantages.  First, within the nuclear network calculation, the smaller 
number of variables results in a smaller matrix to build and solve. 
This speeds the nucleosynthesis calculation by a factor of 2 in the case of 
the \alp\ network, with the potential for much greater speed increases for 
larger networks.  Second, the reduction in the number of nuclear variables 
also reduces the number of equations which must be solved within the 
encompassing hydrodynamic model, a matter of particular concern for models 
with large numbers of zones or grid points, especially multi-dimensional 
studies.  The relative importance of these 2 advantages depends on the
size of the network.  As a rule of thumb, a PPM hydrodynamic code evolving
45 nuclear abundances spends roughly half its time on the network solution
(\cite{Arne98}).  Thus for small networks, like the \alp\ network, the
reduction in necessary number of hydrodynamic equations is the paramount
advantage.

In an effort to streamline the computation necessary for advanced nuclear
burning stages, we have combined a smaller \alp\ network (for calculating 
the nuclear evolution from helium burning through oxygen burning) with this 
QSE based method.  The resulting combined network, which we have dubbed 
\alp 7, can be used in place of the full 14 element \alp\ network from 
helium burning through to NSE without significant errors in energy generation 
or nucleosynthesis. Such a combination offers the same reductions in the 
computational cost discussed above for silicon burning for all burning 
stages beyond H burning.    In a future paper \cite{HiFT99}, we will apply 
the method of QSE-reduction to larger nuclear networks, where the potential 
for improvement in speed and size is even greater.

\acknowledgements

The authors would like to thank P. H\"oflich, K. Nomoto, E. Oran and D. 
Arnett for fruitful discussions.  This research was supported in part by 
NASA Grant NAG5-2888 and NSF Grant AST 9528110.  FKT was supported in part 
by Swiss Nationalfonds grant 20-47252.96

\def\canjphys{{Can. J. Phys.} }

\end{document}